# Full-Duplex Joint Sensing for Opportunistic Access in Spectrum-Heterogeneous Cognitive Radio Networks

Peng Liu, *Student Member, IEEE*, Wangdong Qi, *Member, IEEE*, Li Wei, En Yuan and Bing Xu

*Abstract*—In cognitive radio networks (CRN), secondary users (SUs) can share spectrum with licensed primary users (PUs). Because an SU receiver (SU-Rx) does not always share the same view of spectrum availability as the corresponding SU transmitter (SU-Tx), spectrum sensing conducted only by an SU transmitter tends to be overly sensitive to guarantee safe spectrum access at the price of SU inefficiency. In this letter, we propose a joint spectrum sensing mechanism, named Full-Duplex Joint Sensing (FJDS), to relax sensitivity of SU detection and improve SU throughput. FDJS employs instantaneous feedback enabled by in-band full duplex communication to facilitate the sharing of spectral information from SU-Rx to SU-Tx. The joint detection problem in FDJS is modeled as non-linear optimization and solved by a binary searching algorithm. Simulation results show that FDJS could improve SU throughput as well as meeting PU interruption constraints in a wide range of parameter settings.

## I. INTRODUCTION

IN cognitive radio networks (CRN), unlicensed secondary users (SUs) could utilize the licensed but temporary free spectrum on a "DO NO HARM" basis[1]. Spectrum sensing is an important approach for an SU to monitor the activity of a primary user (PU) and find the available spectrum[2]. It is commonly assumed in CRN research that SUs share the same spectral view; therefore, spectrum sensing is usually conducted only by SU transmitters (SU-Tx). However, recent on-site surveys have shown that PU spectrum occupancy would change dramatically in the scale of hundreds of meters[3]. Since an SU-Tx and its corresponding receiver (SU-Rx) may be tens of kilometers apart[4], there would be many occurrences of inconsistent views of spectrum availability among SUs, which could lead to unsafe spectrum access[3,16].

One conservative approach to tackle this issue is to make SU-Tx so sensitive that a PU's signal can be detected wherever SU-Rx is located. This method has been adopted by the IEEE 802.22 proposal[4]. However, performance deterioration of CRN can result from such an overly sensitive detection strategy[5].

Another approach is to deploy a sensor network to map spatial distribution of PU signals[6,7]. But dedicated sensor networks are not always feasible due to deployment constraints.

Manuscript received March 16th, 2016, revised in July, 2016.
This research was supported by the National Science Foundation of China (61273047 and 61071115) and by the Natural Science Foundation of Jiangsu Province (BK20130068).
The authors are with the ******University (e-mail: herolp@gmail.com).
Digital Object Identifier 10.1109/LCOMM.2016.**.****.

Joint spectrum sensing in [16] eliminates the necessity of a dedicated infrastructure by inferring the spectrum availability of SU-Rx based on historical observations. Unfortunately, spectrum surveys have revealed that more than half (54%) of PU activity patterns are either fast periodic (<1 s) or highly dynamic[8], and therefore it is difficult to infer their access patterns[9].

This letter describes how we incorporate in-band full duplex (IBFD) communication into spectrum sensing and propose a **Full-Duplex Joint Sensing (FDJS)** mechanism. The self-interference suppression (SIS) capability of full duplex radio enables it to transmit and sense, or transmit and receive, simultaneously[10]. With a full duplex channel, SU-Rx in FDJS could feed back the instantaneous spectral information to SU-Tx, where sensing results from both detectors are fused to make a more informed decision on spectrum availability.

Although some research has introduced full duplex capability into spectrum sensing[11-15], all previous studies focus on SU-Tx optimal scheduling of sensing and transmission, following the assumption that SUs share a consistent spectral view. On the other hand, the information fusion between SU-Tx and SU-Rx is not as intuitive as that of cooperative sensing[7]. Specifically, there might be significant differences in SU signal quality, resulting in performance heterogeneity of their detectors. Combining the sensing result from those detectors with poor signal quality would do more harm than good. FDJS tries to solve this issue by adjusting the detection threshold according to the SNR of SU-Tx and SU-Rx. We design a simple but effective method to set the optimal threshold adaptively based on a binary searching algorithm, and present its performance advantage with numerical analysis. Furthermore, simulations are conducted to show how the improvement of spectrum sensing would translate into the increase of CRN throughput.

The rest of the paper is organized as follows. Section II describes a system model. The FDJS mechanism and the joint detection algorithm are discussed in section III. Performance of FDJS is evaluated by numerical analysis and simulations in section IV. Finally, conclusions are listed in section V.

## II. SYSTEM MODEL

### A. Spectrum Occupancy Model

We follow the spectrum occupancy model as described in 802.22. PU-Tx is deployed statically and its signal covers a disc area of radius R, named the keep-out region. If PU-Tx and SU-Tx are active at the same time, we define it as PU disruption as well as CRN transmission failure, regardless of the location of PU-Rx.

*B. Detector Model with Full Duplex Radio*

The goal of SU spectrum sensing is to find out whether a PU is active or absent. Because of residual self-interference in full duplex operation, the two hypotheses involved in detection of a PU can be represented as follows:

$$r(t) = \begin{cases} w(t) + i(t) & H_0: PU\ is\ absent \\ s(t) + w(t) + i(t) & H_1: PU\ is\ active \end{cases} \quad (1)$$

where r(t) is the received signal and s(t), ω(t), and i(t) represent PU signal, noise, and residual self-interference signal respectively. Here, ω(t) is assumed to be a zero-mean i.i.d random Gaussian signal with variance $\sigma_w^2$ and i(t) is assumed to be a zero-mean random signals with variance $\sigma_i^2$.

The performance of a PU detector is usually captured by the receiver operating characteristic (ROC) curve, which is the relationship between the missed detection rate $P_{md}$ and false alarm rate $P_{fa}$. For energy-based detectors with full duplex radio, $P_{fa}$ and $P_{md}$ can be given by the following equations[11]:

$$P_{fa} = Q\left((\gamma - \alpha_i - 1)\sqrt{\frac{N}{2\alpha_i+1}}\right) \quad (2)$$

$$P_{md} = Q\left((\alpha_i + \alpha_s + 1 - \gamma)\sqrt{\frac{N}{2\alpha_i+2\alpha_s+2\alpha_s\cdot\alpha_i+1}}\right) \quad (3)$$

where $Q(*)$ is the Q-function, $\gamma$ is the detection threshold, $N$ is the number of samples, $\alpha_i = \sigma_i^2/\sigma_w^2$ is the ratio between the residual self-interference signal power and noise power, $\alpha_s = \sigma_s^2/\sigma_w^2$ is the ratio between PU signal power and noise power.

Note that $P_{fa}$ and $P_{md}$ have similar expressions for auto-correlation detectors[17]. After eliminating the independent variable $\gamma$, the ROC model can be summarized as follows:

$$P_{fa} = Q(c \cdot Q^{-1}(P_{md}) + k) \quad (4)$$

where $c$ and $k$ are constants related to the number of samples ($N$), signal quality ($\alpha_s$), and residual of self-interference ($\alpha_i$).

*C. Joint Detection Model*

In FDJS, the SU-Rx detection result is sent back to SU-Tx and fused with that of SU-Tx, using the "AND" rule of hard decision. Therefore, link-level spectrum availability is defined as:

$$l^{Link} = l^T \cdot l^R \quad (5)$$

where $l^X$ represents the spectrum availability in the view of X (X may be T for SU-Tx or R for SU-Rx).

Let $m_X$ and $f_X$ represent the missed detection and false alarm probability of X; then the performance of the joint detector could be represented as follows:

$$P_{md} = m_T \cdot m_R \quad (6)$$
$$P_{fa} = 1 - (1 - f_T) \cdot (1 - f_R) \quad (7)$$

## III. FDJS MECHANISM AND ALGORITHM

*A. FDJS Mechanism*

The workflow of FDJS is divided into two stages: 1) half-duplex sensing and query, and 2) full-duplex sensing and communication.

In the first stage, when SU-Tx senses the spectrum and finds that a PU is absent, it queries SU-Rx for confirmation. To avoid interrupting the PU, SU-Tx transmits a direct sequence spread spectrum (DSSS) signal rather than a common coded query message to the target receiver. The low power spectral density of DSSS ensures no interruption to the PU.

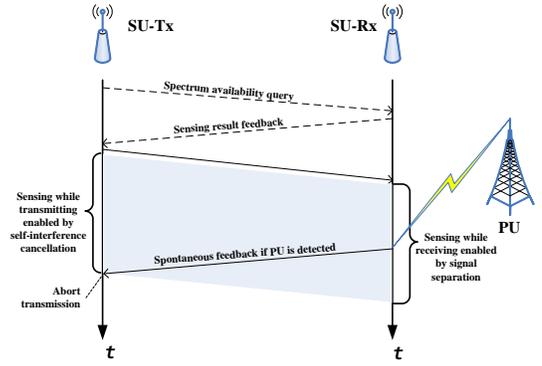

Figure 1. Flow chart of FDJS mechanism.

In the second stage, SU-Tx begins to send frames to SU-Rx once it confirms that the spectrum is also free at SU-Rx. During transmission, it continues to sense the spectrum using full duplex operation. Meanwhile, SU-Rx begins to receive signals from SU-Tx and detects PU activity by searching for a PU's feature signal. Searching is conducted by correlation of the PU feature with the arriving signal. Since searching occurs in parallel with packet decoding, it does not affect the performance of the decoder[18]. Whenever SU-Rx senses a PU's return, it stops ongoing reception and notifies SU-Tx immediately, which forces SU-Tx to abort current transmission and return to stage 1. The flow chart of FDJS is shown in Fig. 1.

*B. Optimal Joint Detection Algorithm*

The goal of FDJS is to find the optimal detection threshold for SU-Tx and SU-Rx. Using the Neyman-Pearson criterion, $P_{md}$ is upper bounded by a constant value, denoted by $b$ here. Then we assign

$$m_T = b^\eta \quad (8)$$
$$m_R = b^{1-\eta} \quad (9)$$

and try to solve the following optimization problem:

$$\text{Min. } P_{fa}$$
$$\text{Sub. to 1) } P_{fa} = f_T + f_R - f_T \cdot f_R$$
$$2)\ f_T = \mathcal{P}(m_T) = Q(c_T \cdot Q^{-1}(b^\eta) + k_T)$$
$$3)\ f_R = \mathcal{R}(m_R) = Q(c_R \cdot Q^{-1}(b^{1-\eta}) + k_R)$$
$$4)\ 0 < \eta < 1$$

where $c_T, k_T, c_R$, and $k_R$ are constants with SU-Tx/Rx. This is a non-linear optimization problem and the optimal solution $\eta^*$ depends on the performance of detectors at SU-Tx and SU-Rx. We assume that detectors at SU-Tx and SU-Rx share the same parameters of $N$ and $\alpha_i$. If $\alpha_s^T = \alpha_s^R$, then $c_T = c_R$ and $k_T = k_R$, and we have $\eta^* = 0.5$, $m_T = m_R = \sqrt{b}$, which is just the case in cooperative sensing. However, SUs might be far away from each other in CRN, and the heterogeneity in their detectors must be considered in optimizing the joint detection. In the rest part of this section, we describe how FDJS finds $\eta^*$ in the case of $\alpha_s^T > \alpha_s^R$, while the deduction in the reverse case is similar.

The derivative of the objective function is as follows:

$$\mathcal{G}(\eta) = \frac{\partial P_{fa}}{\partial \eta} = lnb \cdot \left[ b^\eta \cdot \mathcal{P}'_X(b^\eta) \cdot (1 - \mathcal{R}(b^{1-\eta})) - b^{1-\eta} \cdot \mathcal{R}'_X(b^{1-\eta}) \cdot (1 - \mathcal{P}(b^\eta)) \right] \quad (10)$$

We try to unfold the properties of $P_{fa}(\eta)$ by analyzing its derivative $\mathcal{G}(\eta)$. Substituting $\eta$ in (10) to be 0.5, we have

$$\mathcal{G}(\eta)|_{\eta=0.5} = lnb \cdot \sqrt{b} \cdot \left[ \mathcal{P}'_X(b^{0.5}) \cdot (1 - \mathcal{R}(b^{0.5})) - \mathcal{R}'_X(b^{0.5}) \cdot (1 - \mathcal{P}(b^{0.5})) \right]$$

Under the condition that $\alpha_s^T > \alpha_s^R$ and $\alpha_i^T = \alpha_i^R$, the ROC curve of SU-Tx would always be beneath that of SU-Rx. Consequently, we have

$$0 < \mathcal{P}(x) < \mathcal{R}(x)$$
$$\mathcal{R}'_X(x) < \mathcal{P}'_X(x) < 0$$

and therefore,

$$\mathcal{G}(\eta)|_{\eta=0.5} < 0.$$

Similarly, we have the following conclusions:

$$\mathcal{G}(\eta)|_{\eta=0^+} < 0$$

and

$$\mathcal{G}(\eta)|_{\eta \to 1^-} > 0$$

Extensive numerical simulations have also shown that:

a) $\mathcal{G}(\eta)$ is always minus within the range of (0, 0.5], indicating that $P_{fa}(\eta)$ is a monotone decreasing function in this range. This coincides with the intuition that the higher the SNR of SU-Tx/Rx, the more important role it should play in PU detection, and therefore, $\eta^*$ should never be within (0, 0.5] when $\alpha_s^{Tx} > \alpha_s^{Rx}$.

b) $P_{fa}(\eta)$ is a convex function within the range of (0.5, 1) and its minimum value can be uniquely obtained at the local extreme point where $\mathcal{G}(\eta) = 0$.

Inspired by the above properties, we propose a binary searching based algorithm to find the optimal weight $\eta^*$. The process of this algorithm is shown in TABLE 1.

## IV. PERFORMANCE EVALUATION

In performance evaluation, we mainly concern ourselves with two questions: 1) would the feedback from SU-Rx always improve the performance of joint detection? And, if so, 2) what impact does it make on the throughput of CRN? To answer the first question, we conduct numerical simulations of the detector's performance with/without joint detection. To answer the second question, Monte Carlo simulations are conducted to show how CRN's throughput would improve when cognitive radios are deployed around the edge of a PU's occupancy area ($R = 150$ km).

In simulations, a PU's signal attenuation is modeled as $E_t \cdot r^{-\beta}$. Unless otherwise stated, the parameters are as follows: $E_t = 90$ dBm as that of the DTV transmitter and $\beta = 3.6$ such that the average RSS of PU at the keep-out radius is -96 dBm.

A continuous time Markov chain model is utilized to model the PU's activity pattern. The PU has two states: active (ON) and absent (OFF). The average time of the ON state is $1/\lambda$ while the average time of the OFF state is $1/\mu$. As to the SU traffic pattern, only a saturated link is considered, that is, SU-Tx always has packets to send. The residual self-interference is set to be within -90 to -80 dBm, which has been achieved in

TABLE 1. ALGORITHM TO FIND THE OPTIMAL WEIGHT

Input:
$\alpha_s^T$: SNR of SU-Tx;
$\alpha_s^R$: SNR of SU-Rx;
b: the upper bound of missed detection rate;
N: number of samples during sensing period.
Output:
$\eta^*$: the optimal weight of SU-Tx with joint sensing detector.

| | |
|---|---|
| 1 | if $\alpha_s^T = \alpha_s^R$ |
| 2 | $\eta^*$=0.5; return |
| 3 | else if $\alpha_s^T > \alpha_s^R$ |
| 4 | $\eta_s = 0.5$; $\eta_e = 1$   // set the start and end point of search interval |
| 5 | else |
| 6 | $\eta_s = 0$; $\eta_e = 0.5$ |
| 7 | end |
| 8 | while abs($P_{fa}(\eta_s)$- $P_{fa}(\eta_e)$) > ε (ε is convergence threshold) |
| 9 | $\eta_i = \frac{\eta_s + \eta_e}{2}$ |
| 10 | if $P_{fa}(\eta_i)$ has the same sign with $P_{fa}(\eta_s)$ |
| 11 | $\eta_s = \eta_i$ |
| 12 | else |
| 13 | $\eta_e = \eta_i$ |
| 14 | end |
| 15 | end |
| 16 | $\eta^* = \eta_s$ |

the prototype demo[3].

### A. Performance of Joint Detection

Since the missed detection rate is upper bounded in the Neyman-Pearson criterion, the performance of the detector counts on its false alarm rate. Here, we draw attention to how the false alarm rate varies with respect to SU locations.

Figure 2 shows the deviation of the false alarm rate of a) a traditional cooperative spectrum sensing (CSS) method[7] and b) FDJS against non-cooperative sensing under the same upper bound of $P_{md}$. Energy detector is used here. The detection performance of CSS depends on the location of SU-Rx. If SU-Rx is far from a PU, its feedback would "contaminate" the link level detection decision and increase $P_{fa}$ instead, as represented by the red area in Fig. 2(a). On the contrary, the performance of FDJS always outperforms CSS against a single detector in the whole area of interest. We have also conducted numerical simulation with an auto-correlation detector and varied parameter settings, and got similar results. In short, a weighted detection threshold is a necessity if the communicating SU pair is not located close enough.

### B. Throughput of SUs

The throughput of SUs is dependent on joint detection performance as well as the PU's activity pattern. In particular, the cycle of the PU's state switch (OFF->ON->OFF) has a significant impact on SU throughput. A fast state switch of the PU would definitely increase the disruption rate of SUs and thus decrease their throughput.

There is another side effect of a fast state switch on the **Inference based Joint Sensing (IJS)** method proposed in [16], where SU-Rx's spectral information cannot be obtained in real time. SU-Tx has to infer the spectral availability of SU-Rx according to its historical observations, and would suffer from high inference error when the PU's state changes rapidly.

The simulation results shown in Fig. 3 depict the throughput of CRN with varying PU switch cycles. We compare the performance of FDJS with IJS and CSS. In both simulations, the shared spectrum bandwidth is 6 MHz, the PU's payload is 0.4, the SNR of the SU communication link is 10 dB, and the residual self-interference is -86 dBm. The locations of SUs are different in the two simulations to reflect the impact of SNR.

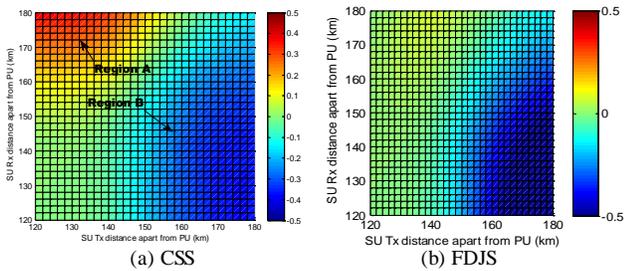

(a) CSS  (b) FDJS

Figure 2. The improvement of detection performance when comparing CSS and FDJS methods. The x- and y-axes are the distances of SU-Tx and SU-Rx apart from a PU, respectively. Those three radios are deployed along one line. In both cases, upper bound is set to be 0.1, and the false alarm rate is expected to be low enough. The color of each grid represents how many times the false alarm rate has been decreased when cooperative/joint sensing is used.

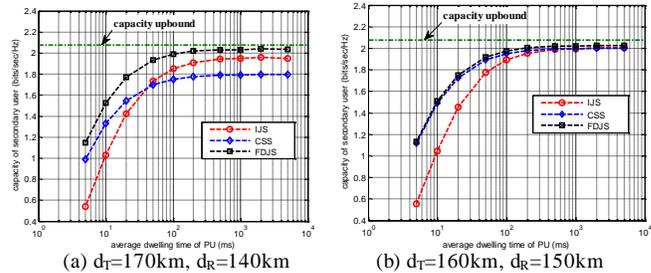

(a) $d_T=170km$, $d_R=140km$  (b) $d_T=160km$, $d_R=150km$

Figure 3. The throughput of CRN with varying PU activity patterns and SU locations. The x-axis is the average time when a PU is active during one switch cycle and the y-axis is the achieved capacity of CRN. The PU's payload is fixed to be 0.4 and thus the average dwelling time is proportional to an ON/OFF switch cycle. The SU radios and PU are assumed to be in one line, therefore, the communicating SU radios are 30 km away from each other in the left figure and 10 km in the right figure. The closer the SU radios, the less different their SNRs. Comparing the two figures, we can see how those detection methods are sensitive to the SU's SNR difference.

The common message delivered by both figures is that the throughput of CRN increases dramatically as a PU's switch cycle enlarges, which is consistent with the qualitative analysis. However, IJS is sensitive to a PU's switch cycle and CSS is sensitive to an SU's location. When a PU's switch cycle decreases, the inference error of IJS plays a major role, making it underperform against CSS. Otherwise, if SUs are far from each other, the large false alarm rate of CSS plays a major role, making it underperform against IJS. FDJS always outperforms IJS and CSS in the whole range of simulation parameters.

## V. Conclusions

Recent spectrum surveys have revealed that there may be many occurrences where SUs have inconsistent views of spectral occupancy in real world deployment. Spectrum sensing conducted only by SU-Tx tries to guarantee safe spectrum access with overly sensitive strategy at the cost of CRN throughput, whereas the sensing fusion between SUs can relax the detection threshold. In-band full duplex communication provides a ready approach for SU-Tx to obtain and fuse SU-Rx's instantaneous spectral information. In this paper, a full-duplex joint sensing mechanism named FDJS is proposed. Further, we designed a simple but effective joint detection algorithm to optimize detection threshold and improve the throughput of CRN. Numerical analysis and simulations with both an energy detector and an auto-correlation detector have shown that FDJS is superior to current methods, including CSS and IJS, within a wide range of parameter settings.